\documentclass[12pt]{iopart}

\usepackage{graphicx}

\newcommand{\pra}{Physical Review A}
\newcommand{\aap}{A\&A}
\newcommand{\apj}{ApJ}
\newcommand{\jgr}{J. Geophysical Research}

\usepackage{iopams}  
\begin{document}

\title[]{Growth rates of the Weibel and tearing mode instabilities in
  a relativistic pair plasma}

\author{J P\'etri and J G Kirk}

\address{Max-Planck-Institut f\"ur Kernphysik, Saupfercheckweg 1,
  69117 Heidelberg, Germany} \ead{j.petri@mpi-hd.mpg.de}

\begin{abstract}
  We present an algorithm for solving the linear dispersion relation
  in an inhomogeneous, magnetised, relativistic plasma. The method is
  a generalisation of a previously reported algorithm that was limited
  to the homogeneous case.  The extension involves projecting the
  spatial dependence of the perturbations onto a set of basis
  functions that satisfy the boundary conditions (spectral Galerkin
  method).  To test this algorithm in the homogeneous case, we derive
  an analytical expression for the growth rate of the Weibel
  instability for a relativistic Maxwellian distribution and compare
  it with the numerical results. In the inhomogeneous case, we present
  solutions of the dispersion relation for the relativistic tearing
  mode, making no assumption about the thickness of the current sheet,
  and check the numerical method against the analytical expression.
 \end{abstract}

\pacs{52.27.Ny; 52.35.-g; 95.30.Qd}
\submitto{\PPCF}

\maketitle

\section{Introduction}

Dissipation of the energy carried by relativistic plasma outflows is
important for the physics of pulsar winds and gamma-ray bursts (for
recent reviews see \cite{2007astro.ph..3116K} and
\cite{2005RvMP...76.1143P}).  In these objects, the plasma is probably
composed of electrons, positrons and protons. As well as being in
relativistic bulk motion with respect to the observer, the random
thermal energy of the plasma may also be relativistic, i.e.,
comparable to the rest mass energy of the constituent particles.  In
this paper we concentrate on two instabilities: these are the
two-stream or Weibel instability~\cite{1959PhRvL...2...83W} and the
tearing modes in a relativistic neutral pair plasma current sheet,
thought to play a role in the formation process of relativistic shocks
\cite{2007arXiv0706.3126S} and the dissipation of magnetic energy in
pulsar winds \cite{2005PPCF...47B.719K}. They are investigated by
generalising and extending to the inhomogeneous case the method
presented in \cite{2007PPCF...49..297P}. Motivated primarily by the
need for code verification, we have derived some analytical
expressions for the linear growth rates of these instabilities.

The Weibel instability is very important in astrophysical processes
because it is able to generate a magnetic field by extracting the free
energy from an anisotropic momentum distribution in an unmagnetised
plasma or from the kinetic drift energy. There is an extensive
literature on the Weibel instability: general conditions for the
existence of the relativistic Weibel instability for arbitrary
distribution functions are discussed in~\cite{2006PhPl...13b2107S},
and wave propagation in counter-streaming magnetised nonrelativistic
Maxwellian plasmas are studied
in~\cite{2005PhPl...12.2901T,2006PhPl...13f2901T}.  Dispersion curves
have been found in some special cases such as, for example, the fully
relativistic bi-Maxwellian distribution function,
(Yoon~\cite{1989PhFlB...1.1336Y}), and the water-bag distribution
function, in which case closed-form analytical expressions can be
derived not only for the Weibel
instability~(Yoon~\cite{1987PhRvA..35.2718Y}), but also for the
cyclotron maser and whistler instabilities
(Yoon~\cite{1987PhRvA..35.2619Y}). However, finding an analytical
expression for the dispersion relation for a given equilibrium
distribution function is a complicated or even impossible task.  It
involves a four-dimensional integration (3D in momentum space and 1D
in time) of the equilibrium distribution function which is difficult
to perform in closed form.  For this reason, the water-bag
distribution is the preferred profile to analyse magnetic field
generation in fast ignitor scenarios, (Silva et
al.~\cite{2002PhPl....9.2458S}) and in relativistic shocks, (Wiersma
and Achterberg~\cite{2004A&A...428..365W}, Lyubarsky and
Eichler~\cite{2006ApJ...647.1250L}). The Weibel instability in a
magnetised electron-positron pair plasma has been investigated by Yang
et al~\cite{1993PhFlB...5.3369Y} using two model distributions: the
water bag, and one with a power-law dependence at high energy. A
general covariant description has been formulated by
Melrose~\cite{1982AuJPh..35...41M} and by
Schlickeiser~\cite{2004PhPl...11g5532S}. In the present work, we focus
on equilibrium configurations given by a relativistic Maxwellian
distribution function, which allows one to reduce the four-dimensional
integral to a simple one-dimensional integral, as we demonstrate in
Section~\ref{sec:weibel_analytical}. The growth rates are then found
by solving this equation using a single numerical quadrature, and are
compared to the results found using our extended algorithm in
Section~\ref{sec:weibel_numerical}.

In the inhomogeneous case, the stability properties of a
nonrelativistic Harris current sheet also have a substantial
literature, with notable recent studies by
Daughton~\cite{1999PhPl....6.1329D} and Silin et
al.~\cite{2002PhPl....9.1104S}.  In the relativistic case, the tearing
mode instability has been investigated by Zelenyi \&
Krasnoselskikh~\cite{1979SvA....23..460Z}, by integrating first order
perturbations of the relativistic Maxwellian distribution function
along approximate, straight-line particle trajectories, in the thick
layer limit (in which the Larmor radius is much smaller than the
thickness of the current sheet).  In
Section~\ref{sec:tearing_analytical} we lift these restrictions to
present new results for the tearing mode instability in a neutral
current sheet of arbitrary temperature and thickness, and compare
these with the results found using the generalised algorithm in
Section~\ref{sec:tearing_numerical}. In this work, no assumption is
made about the thickness of the current sheet, and the particle
trajectories are found numerically in the background magnetic field.

The numerical method, which is an extension of our previous algorithm,
\cite{2007PPCF...49..297P}, that computes the linear dispersion
relation of waves within a Vlasov-Maxwell description, is described in
Section~\ref{sec:algorithm}. It is based on the approach of
Daughton~\cite{1999PhPl....6.1329D} for non relativistic Maxwellians,
and involves explicit time integration of particle orbits along the
unperturbed trajectories. We modify and extend our former code to
include inhomogeneities in the plasma equilibrium configuration.
Moreover, we generalise it to a fully relativistic approach, i.e., one
that allows for relativistic temperatures as well as relativistic
drift speeds.

\section{Analytical treatment of the instabilities}
\label{sec:analytical}
\subsection{The Vlasov-Maxwell equations}
\label{sec:vlasov-maxwell}

For convenience, we first recall the full set of Vlasov-Maxwell equations
governing the non-linear time evolution of the pair plasma.

We introduce the standard electromagnetic scalar and vector
potentials~$(\phi,\vec{A})$, related to the electromagnetic
field~$(\vec{E}, \vec{B})$ by~:
\begin{eqnarray}
  \label{eq:Potentiel}
  \vec{E} & = & - \vec{\nabla} \phi - \frac{\partial\vec{A}}{\partial t} \\
  \vec{B} & = & \vec{\nabla} \wedge \vec{A}
\end{eqnarray}
We employ the Lorenz gauge condition by imposing
\begin{equation}
  \label{eq:Jauge}
  \mathrm{div}\, \vec{A} + \varepsilon_0 \, \mu_0 \, \frac{\partial\phi}{\partial t} = 0
\end{equation}
with $\varepsilon_0 \, \mu_0 \, c^2 = 1$ and $c$ the speed of light.
The relation between potentials and sources then reads~: \numparts
\begin{eqnarray}
  \label{eq:OndePhi}
  \Delta \phi - \frac{1}{c^2} \, \frac{\partial^2\phi}{\partial t^2} +
  \frac{\rho}{\varepsilon_0} & = & 0 \\
  \label{eq:OndeA}
  \Delta \vec{A} - \frac{1}{c^2} \, \frac{\partial^2\vec{A}}{\partial t^2} + 
  \mu_0 \, \vec{j} & = & 0
\end{eqnarray}
\endnumparts The source terms represented by the charge~$\rho$ and
current~$\vec{j}$ densities, are expressed in terms of the
distribution functions of each species,~$f_s$, by~: \numparts
\begin{eqnarray}
  \label{eq:SourceRho}
  \rho(\vec{r},t) & = & \sum_s q_s \, \int\!\!\!\int\!\!\!\int 
  f_s(\vec{r},\vec{p},t) \, \rmd^3\vec{p} \\
  \label{eq:SourceJ}
  \vec{j}(\vec{r},t) & = & \sum_s q_s \, \int\!\!\!\int\!\!\!\int 
  \frac{\vec{p}}{\gamma\,m_s} \, f_s(\vec{r},\vec{p},t) \, \rmd^3\vec{p}
\end{eqnarray}
\endnumparts where $\gamma = \sqrt{1+\vec{p}\,^2/m_s^2\,c^2}$ is the
Lorentz factor of a particle. We adopted the usual notations, namely
$(t, \vec{r}, \vec{v}, \vec{p}, m_s, q_s)$ for respectively the time,
position, 3-velocity, 3-momentum, mass and charge of a particle of
species~$s$.  The time evolution of the distribution functions~$f_s$
is governed by a relativistic Vlasov equation for each species:
\begin{equation}
  \label{eq:Vlasov}
  \frac{\partial f_s}{\partial t} + \vec{v} \cdot \frac{\partial f_s}{\partial \vec{r}}
  + q_s \, ( \vec{E} + \vec{v} \wedge \vec{B} ) \,
  \cdot \frac{\partial f_s}{\partial \vec{p}} = 0
\end{equation}
The self-consistent non-linear evolution of the plasma is entirely
determined by the set of
equations~(\ref{eq:Potentiel})-(\ref{eq:Vlasov}).

The linear stability properties of the neutral current sheet are
investigated by linearising the set of equations
(\ref{eq:Potentiel})-(\ref{eq:Vlasov}). The procedure has been
described in~\cite{2007PPCF...49..297P}, Sec.~3. These results are now
used to derive linear dispersion relations.

\subsection{Weibel instability}
\label{sec:weibel_analytical}

In the Weibel regime, the electrostatic potential~$\phi$ can be
neglected. Moreover, only the $A_{\rm y}$ component of the vector
potential~$\vec{A}$ comes into play in the eigenvalue problem. Note
also that the Weibel instability corresponds to low-frequency modes
propagating along the $z$-axis such that
\begin{equation}
  \label{eq:LowFreq}
  ||\omega|| \ll k_{\rm z} \, c  
\end{equation}
with $k_{\rm y}=0$, $\vec{k}=(0, k_{\rm y}, k_{\rm z})$ being the
wavenumber and $\omega$ the corresponding eigenfrequency.

Specialising the general derivation of the perturbed charge density
and current density presented in Sect.~3 of \cite{2007PPCF...49..297P}
to the Weibel instability, setting $\phi = A_{\rm x} = A_{\rm z} =
k_{\rm y} = 0$, the dispersion relation reads
\begin{eqnarray}
  \label{eq:Weibel1}
  \frac{\omega^2}{c^2} - k_{\rm z}^2 + 
  \frac{2 \, \Gamma_s^2 \, \beta_s^2 \, \omega_{{\rm p}s}^2}{c^2 \, \Theta_s} \, +
  \sum_s i \, \omega \, \int\!\!\!\int\!\!\!\int 
  \frac{f_{0s} \, p_{\rm y}}{\gamma \, m_s^2} \, & \times & \nonumber \\
  \times \int_{-\infty}^0 p_{\rm y}' \, 
  {\rm exp}[i \, ( k_{\rm z} \, p_{\rm z} / m_s - \gamma \, \omega ) \, \tau' ] \, 
  d\tau' \, d^3\vec{p} = 0 & & 
\end{eqnarray}
The distribution function at equilibrium for each species~''$s$'',
denoted by $f_{0s}(\vec{r},\vec{p})$, is assumed to be a relativistic
Maxwellian with constant drift speed~$\pm U_s$.  The stationary
distribution function for each species reads:
\begin{equation}
  \label{eq:FDDWeibel}
  f_{0s}(\vec{r},\vec{p}) = 
  \frac{N_s}{4 \, \pi \, m_s^3 \, c^3 \, \Theta_s \, K_2(1/\Theta_s)}   
  \, \mathrm{exp}[-\Gamma_s \, ( E - U_s \, p_\mathrm{y} ) / \Theta_s \, m_s \, c^2]
\end{equation}
$N_s$ is the (constant) particle number density of the plasma,
$\omega_{{\rm p}s} = N_s \, q_s^2 / m_s \, \varepsilon_0$ the plasma
frequency, $E = m_s \, c^2 \, \sqrt{1 + \vec{p}\,^2/m_s^2\,c^2}$ the
total energy of a particle, $p_\mathrm{y}$ the y-component of its
momentum, $\Gamma_s = 1 / \sqrt{ 1 - \beta_s^2}$ ($\beta_s = U_s/c$)
the Lorentz factor associated with the drift motion and $K_2$ the
modified Bessel function of the second kind and of order~2. The
temperature of the gas~$T_s$ is conveniently normalised to the rest
mass energy of the leptons such that
\begin{equation}
  \label{eq:Theta}
  \Theta_s = \frac{k_B \, T_s}{m_s \, c^2}
\end{equation}
and $k_B$ is the Boltzmann constant.  The triple integral along the
momentum vector can be performed analytically in the following way.

Because no external background electromagnetic field exists at
equilibrium, the particle trajectories are straight lines. The
integration of the equations of motion leads to
\begin{eqnarray}
  \label{eq:sd}
  \vec{p}\,' & = & \vec{p} = {\rm cst} \\
  \vec{r}\,' & = & \vec{r} + \frac{\vec{p}}{m_s} \, \tau' 
\end{eqnarray}
Therefore time and momentum integration can be inverted. The
dispersion relation thus reduces to
\begin{eqnarray}
  \label{eq:Weibel2}
  \frac{\omega^2}{c^2} - k_{\rm z}^2 + 
  \frac{2 \, \Gamma_s^2 \, \beta_s^2 \, \omega_{{\rm p}s}^2}{c^2 \, \Theta_s} \, +
  \sum_s i \, \omega \, \frac{\Gamma_s \, \omega_{{\rm p}s}^2}{c^4 \, \Theta_s} \, 
  \int_{-\infty}^0 
  \int\!\!\!\int\!\!\!\int \frac{p_{\rm y}^2}{\gamma \, m_s^2} \, \times \\
  \times \, 
  \frac{{\rm exp}( - \Gamma_s \, ( E - U_s \, p_{\rm y}) / \Theta_s \, m_s \, c^2) }
  {4 \, \pi \, m_s^3 \, c^3 \, \Theta_s \, K_2(1/\Theta_s)} \,
  {\rm exp}[i \, ( k_{\rm z} \, p_{\rm z} / m_s - \gamma \, \omega ) \, \tau' ]
  \, d^3\vec{p} \, d\tau' = 0 \nonumber
\end{eqnarray}
The integration along the momentum~$\vec{p}$ can be done analytically
with help on the following formula, see for
instance~\cite{1958PhDT........18T}
\begin{equation}
  \label{eq:Trubnikov1}
  I(A,\vec{\alpha}) \equiv \frac{1}{4\,\pi} \, \int\!\!\!\int\!\!\!\int 
  \frac{{\rm exp}(-A \, \gamma \pm i \, \vec{\alpha} \cdot \vec{p})}
  {\gamma} \, d^3\vec{p} = m_s^3 \, c^3 
  \frac{K_1(\sqrt{A^2 + m_s^2 \, c^2 \, \alpha^2})}
  {\sqrt{A^2 + m_s^2 \, c^2 \, \alpha^2}}
\end{equation}
where the Lorentz factor of a particle is $\gamma = \sqrt{1 +
  \vec{p}\,^2 / m_s^2 \, c^2}$ and $K_1$ is the modified Bessel
function of order~1.  By differentiating twice with respect to the
$y$~component, $\alpha_{\rm y}$, of the vector~$\vec{\alpha}$, we get
\begin{eqnarray}
  \label{eq:Trubnikov2}
  I_{\rm py}(A,\vec{\alpha}) & \equiv & 
  - \frac{\partial^2 I}{\partial\alpha_{\rm y}^2}(A,\vec{\alpha}) = 
  \frac{1}{4\,\pi} \, \int\!\!\!\int\!\!\!\int 
  \frac{p_{\rm y}^2 \, {\rm exp}(-A \, \gamma \pm i \, \vec{\alpha} \cdot \vec{p})}
  {\gamma} \, d^3\vec{p} = \\
  & & m_s^5 \, c^5 \, \left[ \frac{A^2 + m_s^2 \, c^2 \, \alpha^2 - 
      4 \, m_s^2 \, c^2 \, \alpha_{\rm y}^2}{(A^2 + m_s^2 \, c^2 \, \alpha^2)^2}
    \, K_2(\sqrt{A^2 + m_s^2 \, c^2 \, \alpha^2}) - \right. \nonumber \\
  & & \left.
    \frac{ m_s^2 \, c^2 \, \alpha_{\rm y}^2}{(A^2 + m_s^2 \, c^2 \, \alpha^2)^{3/2}} \,
    K_1(\sqrt{A^2 + m_s^2 \, c^2 \, \alpha^2}) \right] \nonumber
\end{eqnarray}
Applying these formulae to our problem, it is convenient to introduce
the following quantities
\begin{eqnarray}
  A(\omega,\tau) & = & \frac{\Gamma_s}{\Theta_s} + i \, \omega \, \tau \\
  \vec{\alpha}(\tau) & = & - i \, \frac{\Gamma_s \, \beta_s}{\Theta_s \, m_s \, c} 
  \, \vec{e}_{\rm y} + \frac{k_{\rm z} \, \tau}{m_s} \, \vec{e}_{\rm z}
\end{eqnarray}
The function~$I_{\rm py}$ depends now on $\omega$ and $\tau$
via~$A(\omega,\tau)$ and $\vec{\alpha}(\tau)$, $I_{\rm
  py}(A(\omega,\tau),\vec{\alpha}(\tau))$, assuming that the
equilibrium quantities such as species temperatures~$\Theta_s$ and
drift speeds~$\beta_s$ are prescribed and therefore constant. In the
remainder of this section, we will denote it for simplicity by~$I_{\rm
  py}(\omega,\tau)$.  The dispersion relation for the Weibel
instability, Eq.~(\ref{eq:Weibel2}) therefore reads
\begin{equation}
  \label{eq:RelDispWeibel}
  \frac{\omega^2}{c^2} - k_{\rm z}^2 + 
  \frac{2 \, \Gamma_s^2 \, \beta_s^2 \, \omega_{{\rm p}s}^2}{c^2 \, \Theta_s} \, +
  \frac{2 \, i \, \omega \, \Gamma_s \, 
    \omega_{{\rm p}s}^2}{c^2 \, \Theta_s^2 \, K_2(1/\Theta_s)} 
  \, \frac{1}{m_s^5 \, c^5} \, \int_{-\infty}^0 I_{\rm py}(\omega,\tau) \, d\tau = 0
\end{equation}
In the non-relativistic limit, $\Theta_s\ll1$, we recover the
dispersion relation Eq.(33) of \cite{2007PPCF...49..297P}.

\subsection{Relativistic neutral current sheet}
\label{sec:Equilibre}

We now turn to the study of the unstable tearing mode 
in a relativistic neutral
current sheet of thickness~$L$ and asymptotic magnetic field
intensity~$B_0$ (far from the current sheet). The plasma is
one-dimensional in the sense that it has only spatial variation in the
$x$ direction. It consists of counter-streaming electrons and positrons
with relativistic temperatures~$T_s$ evolving in a static external
magnetic field aligned with the $z$-axis such
that~\cite{1966PhFl...9..277H}
\begin{equation}
  \label{eq:ChampMag}
  B_{\rm z}(\vec{r}) = B_0 \, \tanh \left( \frac{x}{L} \right)
\end{equation}
We use Cartesian coordinates, denoted by~$\vec{r} = ({x,y,z})$, and
the corresponding basis~$(\vec{e}_\mathrm{x}, \vec{e}_\mathrm{y},
\vec{e}_\mathrm{z})$. In equilibrium, there is no electric field,
$\vec{E}_0 = \vec{0}$ and the charges drift in the $y$~direction at a
relativistic velocity~$U_s$.  The particle number density for each
species is
\begin{equation}
  \label{eq:Densite}
  n_{0s}(\vec{r}) = N_s \, {\rm sech}^2 \left( \frac{x}{L} \right)
\end{equation}
The distribution function at equilibrium for each species~''$s$'',
denoted by $f_{0s}(\vec{r},\vec{p})$, is assumed to be a relativistic
Maxwellian with constant drift speed~$\pm U_s$.  The stationary
distribution function for each species reads:
\begin{equation}
  \label{eq:FDD}
  f_{0s}(\vec{r},\vec{p}) = \frac{n_{0s}(\vec{r})}
  {4 \, \pi \, m_s^3 \, c^3 \, \Theta_s \, K_2(1/\Theta_s)}
  \, \mathrm{exp}[-\Gamma_s \, ( E - U_s \, p_\mathrm{y} ) / \Theta_s \, m_s \, c^2]
\end{equation}
$n_{0s}(\vec{r})$ is the particle number density in the current sheet,
Eq.~(\ref{eq:Densite}), and all other quantities are the same as those
defined for the distribution function in Eq.~(\ref{eq:FDDWeibel}). The
stationary Vlasov-Maxwell equations are satisfied provided that
\begin{eqnarray}
  \label{eq:Pressure}
  4 \, N_s \, \Theta_s \, m_s \, c^2 & = & \frac{B_0^2}{\mu_0} \\
  \Gamma_s \, U_s & = & - 2 \, \frac{\Theta_s \, m_s \, c^2}{q_s \, B_0 \, L}
\end{eqnarray}
Eq.~(\ref{eq:Pressure}) states the balance between gaseous pressure
and magnetic pressure at the centre of the current sheet. It is also
useful to note that the non-relativistic plasma frequency (at the
centre of the current sheet) and cyclotron frequency defined,
respectively, by
\begin{eqnarray}
  \label{eq:FreqPlasma}
  \hat{\omega}_{ps}^2 & = & \frac{N_s \, q_s^2}{m_s \, \varepsilon_0} \\
  \label{eq:FreqCylcotron}
  \hat{\Omega}_{Bs} & = & \frac{q_s \, B_0}{m_s}
\end{eqnarray}
are related by
\begin{equation}
  \label{eq:Freq}
  4 \, \hat{\omega}_{ps}^2 \, \Theta_s = \hat{\Omega}_{Bs}^2
\end{equation}
These frequencies are constant (independent of $x$), and are denoted
by a~$\hat{}$ in order to distinguish them from quantities that depend
on~$x$.

\subsection{The relativistic tearing mode}
\label{sec:tearing_analytical}

A method similar to that used in Section~\ref{sec:weibel_analytical}
can also be applied to the relativistic neutral current sheet.  Using
the equilibrium distribution function presented in Eq.~(\ref{eq:FDD})
for the relativistic Harris current sheet, the eigenvalue equation to
solve reads
\begin{eqnarray}
  \label{eq:Harris1}
  A_{\rm y}''(x) - \left( k_{\rm z}^2 - \frac{\omega^2}{c^2} \right) \, A_{\rm y}(x) +
  \frac{2}{L^2} \, {\rm sech}^2 \left( \frac{x}{L} \right) \, A_{\rm y}(x) & + &
  \nonumber \\
  \frac{2 \, i \, \omega \, \Gamma_s \, 
    \hat{\omega}_{{\rm p}s}^2}{c^2 \, \Theta_s^2 \, K_2(1/\Theta_s)} 
  \, \frac{A_{\rm y}(x)}{m_s^5 \, c^5} \, \int_{-\infty}^0 I_{\rm py}(\omega,\tau) \, d\tau = 0
\end{eqnarray}
where prime~$''$ denotes second derivative with respect to~$x$.
We now follow \cite{1991JGR....9611523P} and introduce the 
variable $t=\tanh(x/L)$,  which transforms Eq.~(\ref{eq:Harris1}) 
into the Legendre equation. From this, one sees that the 
dispersion relation is satisfied 
by purely growing modes, $\textrm{Re}(\omega)=0$, 
the tearing modes, whose growth rate is the solution of 
\begin{equation}
  \label{eq:Harris2}
   \frac{2 \, i \, \omega \, \Gamma_s \, \hat{\omega}_{{\rm p}s}^2 \, L^2}
   {c^2 \, \Theta_s^2 \, K_2(1/\Theta_s)} \, \frac{1}{m_s^5 \, c^5} \, 
   \int_{-\infty}^0 I_{\rm py}(\omega,\tau) \, d\tau = ( k_{\rm z} \, L + 2 ) \, ( k_{\rm z} \, L - 1 )
\end{equation}
where we used the \lq\lq low-frequency\rq\rq\ approximation,
Eq.~(\ref{eq:LowFreq}), to neglect the contribution of the
displacement current.  Expression~(\ref{eq:Harris2}) contains a single
one-dimensional integral. Finding its solutions is computationally
much faster than dealing with the general expression in 4~dimensions
and does not involve any approximation concerning the current sheet
thickness.

In the low-temperature limit, $\Theta_s\ll1$ (non-relativistic case),
the dispersion relation Eq.~(\ref{eq:Harris1}) reduces to its
classical expression given by
\begin{equation}
  \label{eq:Tearing}
  \frac{\omega}{k_{\rm z} \, v_{\rm th}} \, 
  Z \left( \frac{\omega}{k_{\rm z} \, v_{\rm th}} \right) 
  \sum_s \left( 1 + 2 \, \frac{U_s^2}{v_{\rm th}^2} \right) = 
  \frac{c^2}{\hat{\omega}_{{\rm p}s}^2 \, L^2} \, 
  ( k_{\rm z} \, L + 2 ) \, ( k_{\rm z} \, L - 1 )
\end{equation}
where the plasma dispersion function~$Z$ is defined for
$\mathrm{Im}(\zeta)>0$ by
\begin{equation}
  \label{eq:FonctionZ}
  Z(\zeta) = \frac{1}{\sqrt{\pi}} \, \int_{-\infty}^{+\infty}
  \frac{\rme^{-t^2}}{t - \zeta} \, \rmd t
\end{equation}
and analytically continued for $\mathrm{Im}(\zeta)<0$, see for
instance Delcroix and Bers~\cite{1994Delcroix}.

\section{The algorithm}
\label{sec:algorithm}

We first recall the general linear eigenvalue system to be solved, as
presented in~\cite{2007PPCF...49..297P} and then present the extended
algorithm for inhomogeneous and magnetised plasmas.

\subsection{The eigenvalue system}

The eigenvalue system is found by solving the equations for the
electromagnetic potential determined according to the source
distribution given by~(\ref{eq:SourceRho}) and (\ref{eq:SourceJ}).
Inserting the latter expressions into~(\ref{eq:OndePhi})
and~(\ref{eq:OndeA}), the eigenvalue system reads~: \numparts
\begin{eqnarray}
  \label{eq:SysPropPhiMono}
  \phi''(x) - \left( \vec{k}\,^2 - \frac{\omega^2}{c^2} \right) \, \phi(x) +
  \frac{\rho(x)}{\varepsilon_0} & = & 0 \\
  \label{eq:SysPropAMono}
  \vec{A}''(x) - \left( \vec{k}\,^2 - \frac{\omega^2}{c^2} \right) \, \vec{A}(x) + 
  \mu_0 \, \vec{j}(x) & = & 0
\end{eqnarray}
\endnumparts For homogeneous plasmas, the system reduces to a
$4\times4$ matrix, which is easily solved, see
\cite{2007PPCF...49..297P}. In the inhomogeneous case, the
perturbations should vanish asymptotically, when $x = \pm \infty$. To
solve this inhomogeneous eigenvalue problem, it is convenient to
expand the unknown quantities on a set of basis functions, each
function individually satisfying the required boundary conditions.
This method is known as a spectral Galerkin method, similar to Fourier
decomposition for periodic boundary conditions. We will refer to it as
a spectral decomposition as described below.

\subsection{Spectral decomposition}

The perturbations are projected on an orthonormal set of basis
functions, $\mathcal{F}_n$. Because of the required boundary
conditions, namely $\phi(x=\pm\infty)=0$, $\vec{A}(x=\pm\infty)=0$,
the Hermite functions are a convenient set onto which to expand the
perturbations, \cite{1999PhPl....6.1329D}. They are given by
\begin{equation}
  \label{eq:HermiteF}
  \mathcal{F}_n(x) = \frac{H_n(x)}{\sqrt{2^n \, n! \, \sqrt{\pi}}} \, \exp[-x^2/2]
\end{equation}
where $H_n$ are the Hermite polynomials, \cite{1953mtp..book.....M}.
We introduce the projection operator of a function~$g$ on a basis
function~$\mathcal{F}_n$ defined as
\begin{equation}
  \label{eq:projecteur}
  <g|\mathcal{F}_n> \equiv \int_{-\infty}^{+\infty} g(x) \, \mathcal{F}_n(x) \, dx
\end{equation}
The basis functions themself satisfy the orthonormality relation
\begin{equation}
  \label{eq:Normalite}
   <\mathcal{F}_n|\mathcal{F}_k> = \delta_{nk}
\end{equation}
where $\delta_{nk}$ is the kronecker symbol. The matrix representation
of the second derivative, useful for projection of
Eq.~(\ref{eq:SysPropPhiMono}) and (\ref{eq:SysPropAMono}), is then
represented by
\begin{eqnarray}
  \label{eq:MatriceDeriv}
  \mathcal{D}_{nk} & \equiv & \int_{-\infty}^{+\infty} \mathcal{F}_n(x) \, 
  \mathcal{F}_k''(x) \, dx \nonumber \\
  & = & \frac{\sqrt{( n + 1 ) \, ( n + 2 )}}{2} \, \delta_{n,k+2} - 
  \frac{2\,n+1}{2} \, \delta_{nk} + 
  \frac{\sqrt{ n \, ( n - 1 )}}{2} \, \delta_{n,k-2}
\end{eqnarray}
For convenience, we introduce the unknown four dimensional vector
$\vec{\Psi} = (\phi, \vec{A})$. Using $N$~terms in the expansion
of~$\vec{\Psi}$, it is written as
\begin{equation}
  \label{eq:Her}
  \vec{\Psi}(x) = \sum_{k=0}^{N-1} \vec{\Psi}_k \, \mathcal{F}_k(x)
\end{equation}
Inserting this expression in Eq.~(\ref{eq:SysPropPhiMono})
and~(\ref{eq:SysPropAMono}) and projecting them on each basis
function~$\mathcal{F}_n$, the system reduces to a matrix equation of
dimension $4N\times4N$ for the $N$~unknown vectors~$\vec{\Psi}_k$
\begin{equation}
  \label{eq:SysProp}
  M(\omega,\vec{k}) \cdot \vec{\Psi} = 0
\end{equation}
More explicitly, let $C_k$ be the $4N$ unknowns. We choose to order
the unknowns in the following way, $\vec{\Psi}_k = (C_k, C_{k+N},
C_{k+2N}, C_{k+3N})$. The expansion Eq.~(\ref{eq:Her}) is done such that
\numparts
\begin{eqnarray}
  \phi(x) & = & \sum_{k=0}^{N-1} C_k \, \mathcal{F}_k(x) 
\label{sum1}\\
  A_{\rm x}(x) & = & \sum_{k=0}^{N-1} C_{k+N} \, \mathcal{F}_k(x) 
\label{sum2}\\
  A_{\rm y}(x) & = & \sum_{k=0}^{N-1} C_{k+2N} \, \mathcal{F}_k(x) 
\label{sum3}\\
  A_{\rm z}(x) & = & \sum_{k=0}^{N-1} C_{k+3N} \, \mathcal{F}_k(x)
\label{sum4}
\end{eqnarray}
\endnumparts Eq.~(\ref{eq:SysProp}) is a {\em non-linear} eigenvalue
problem for the matrix~$M$ with eigenvector $\vec{\Psi}$ and
eigenvalue~$\omega$.  The method of solution has been described in
\cite{2007PPCF...49..297P}.  However, the simultaneous search of the
eigenvalues and the eigenvectors is very time-consuming because the
root finding takes place in a $8N+2$ dimension parameter space ($4N+1$
complex numbers to find, the eigenvalue~$\omega$ and the $C_k$).
Therefore, we decide only to look for the vanishing of the matrix
determinant,~${\rm det}\, M(\omega,\vec{k})=0$. In this paper, we only
present the capabilities of our algorithm and do not discuss in detail
the physics of the unstable perturbations.  For this purpose, it is
sufficient to look only for the vanishing of the determinant of the
matrix~$M$. Nevertheless, we present an example of an eigenfunction in
Section~\ref{sec:tearing_numerical}, where we examine the convergence
properties of the summations in Eqs.~(\ref{sum1}--\ref{sum4}).

\section{Results}
\label{sec:results}

\subsection{Weibel instability}
\label{sec:weibel_numerical}

\begin{figure}[htbp]
  \centering
  \includegraphics[scale=0.8]{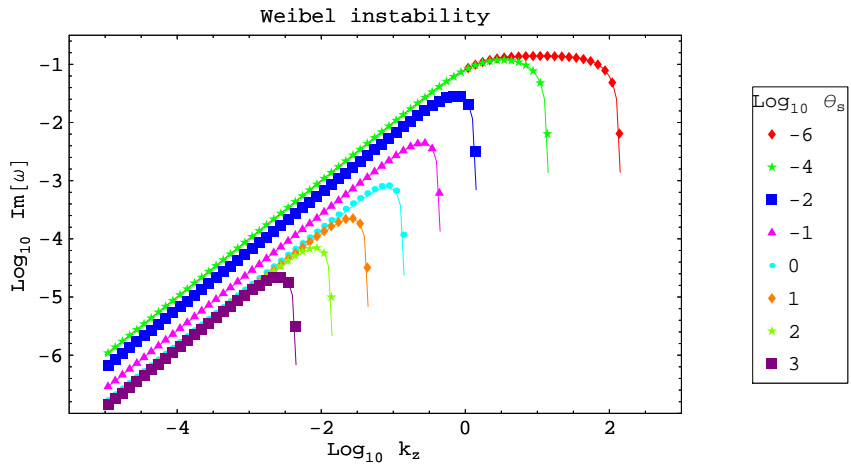}
   \caption{Growth rates of the Weibel instability for different 
     temperatures~$\Theta_s$, see legend, and a given drift
     speed~$\beta_s = 0.1$.  Points are results from the algorithm and
     the solid lines are from the numerical solution to the dispersion
     relation, Eq.~(\ref{eq:RelDispWeibel}).}
  \label{fig:Weibel1}
\end{figure}

As a first check, we compute the dispersion relation for the Weibel
instability for arbitrary plasma temperature~$\Theta_s$. The fully
relativistic algorithm is checked against the numerical solution to
the exact analytical dispersion relation,
Eq.~(\ref{eq:RelDispWeibel}).  The evolution of the growth rates with
increasing plasma temperature at a given drift speed is shown in
fig~\ref{fig:Weibel1}. In our previous work,
\cite{2007PPCF...49..297P}, the non-relativistic case was solved with
a non-relativistic code whereas in the present work, our algorithm
deals with arbitrary temperatures (especially in the
limit~$\Theta_s \ll 1$)%
\footnote{However, this limit needs special
  care because of the exponentially decreasing of the Bessel
  function~$K_2$ in the expressions Eq.~(\ref{eq:FDD}), implying
  numerical under- and overflow problems.}. 
In particular, the results
of the non-relativistic Weibel instability are recovered with the
fully relativistic algorithm.  Inspecting
fig~\ref{fig:Weibel1}, the solutions of our algorithm (denoted by
symbols) are in perfect agreement with the growth rates given by
numerically solving Eq.~(\ref{eq:RelDispWeibel}) (solid lines), for
arbitrary temperature.  Due to the spread in momentum present at
finite temperature, the instability is suppressed for large
wavenumbers~$k_\mathrm{z}$. More precisely, when the temperature
increases, the largest unstable wavenumber decreases. For a given
drift speed, the growth rates decrease with increasing temperature.

\begin{figure}[htbp]
  \centering
  \includegraphics[scale=0.9]{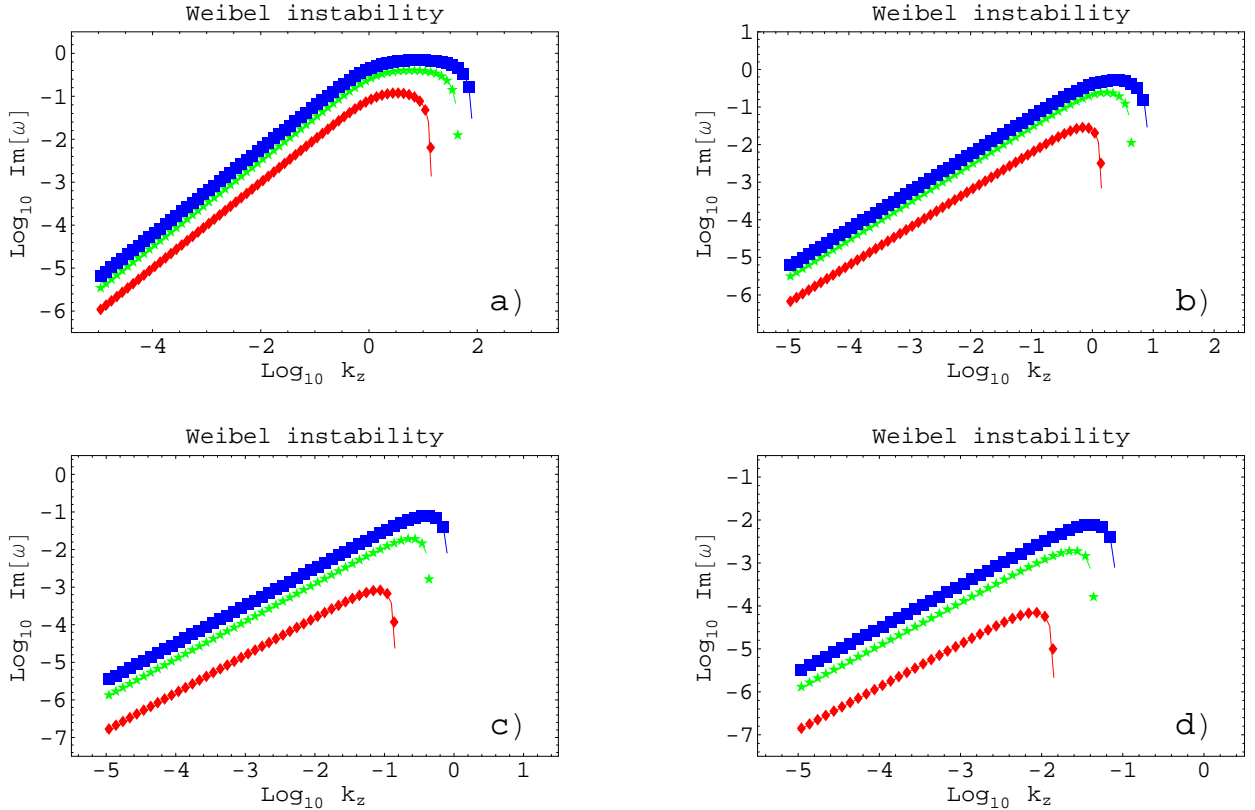} 
   \caption{Growth rates of the Weibel instability 
     for different drift speeds~$\beta_s=0.1/0.3/0.5$, respectively
     red diamonds, green stars and blue squares, and a given
     temperature, $\Theta_s = 10^{-4}$ in fig a), $\Theta_s = 10^{-2}$
     in fig b), $\Theta_s = 1$ in fig c), $\Theta_s = 10^2$ in fig d).
     Points are from our algorithm whereas the solid lines are from
     the numerical solution to the dispersion relation,
     Eq.~(\ref{eq:RelDispWeibel}).}
  \label{fig:Weibel2}
\end{figure}

We also investigate the effect of the drift speed on the growth rates
for a fixed temperature. Results are shown in the non-relativistic
limit $\Theta_s=10^{-4},10^{-2}$, the mildly relativistic case
$\Theta_s=1$, and the ultra-relativistic case $\Theta_s=10^2$, for
increasing drift speed $\beta_s=0.1/0.3/0.5$, respectively red
diamonds, green stars and blue squares, fig.~\ref{fig:Weibel2}.

The characteristic cut-off wavenumber, $k_{\rm cut}$, can be estimated
by setting $\omega=0$ in the dispersion relation
Eq.~(\ref{eq:RelDispWeibel}). Doing so we find
\begin{equation}
  \label{eq:kmax}
  \frac{k_{\rm cut} \, c}{\omega_{{\rm p} s}} =
  \Gamma_s \, \beta_s  \, \sqrt{\frac{2}{\Theta_s}}  
\end{equation}
Moreover, in the ultrarelativistic limit, for high
temperature~$\Theta_s \gg 1$ but small drift speeds~$\beta_s \ll 1$,
the growth rates in the small wavenumber limit are simply given by
\begin{equation}
  \label{eq:GammaLimitRelat}
  \gamma_{\rm rel} = \frac{4}{\pi} \, \beta_s^2 \, k_{\rm z} \, c
\end{equation}
This result is in agreement with the curves shown in
fig.~\ref{fig:Weibel1}.  Note that these growth rates become
independent of the temperature, as in the non-relativistic limit (low
temperature), where
\begin{equation}
  \label{eq:GammaLimit}
  \gamma_{\rm clas} = \beta_s \, k_{\rm z} \, c
\end{equation}
Therefore, in the ultrarelativistic temperature limit, the growth
rates are reduced by a factor $4\,\beta_s/\pi \approx 1.27 \, \beta_s$.

\subsection{Tearing mode}
\label{sec:tearing_numerical}
\begin{figure}[htbp]
  \centering
  \includegraphics[scale=1]{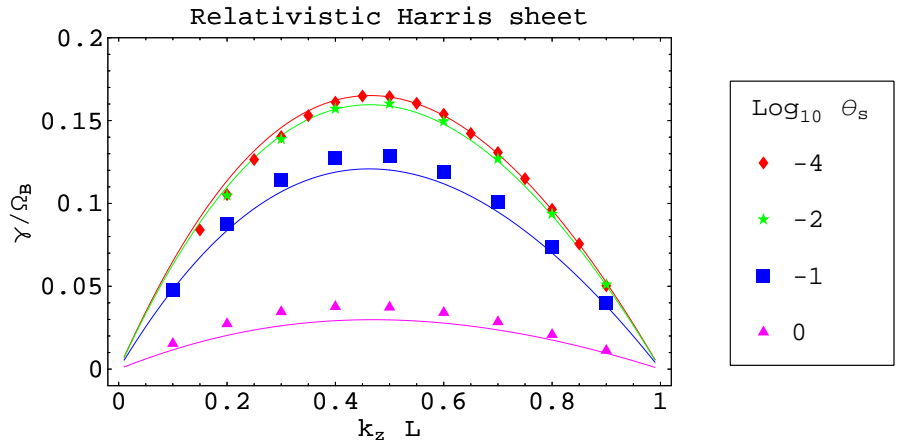}
  \caption{Growth rates of the tearing modes in the Harris current sheet
    for $\rho=L$.  Points are from our numerical algorithm and the
    solid line represents the analytical approximation in
    Eq.~(\ref{eq:Harris2}).}
  \label{fig:Tearing}
\end{figure}

\begin{figure}[htbp]
  \centering \includegraphics[scale=1]{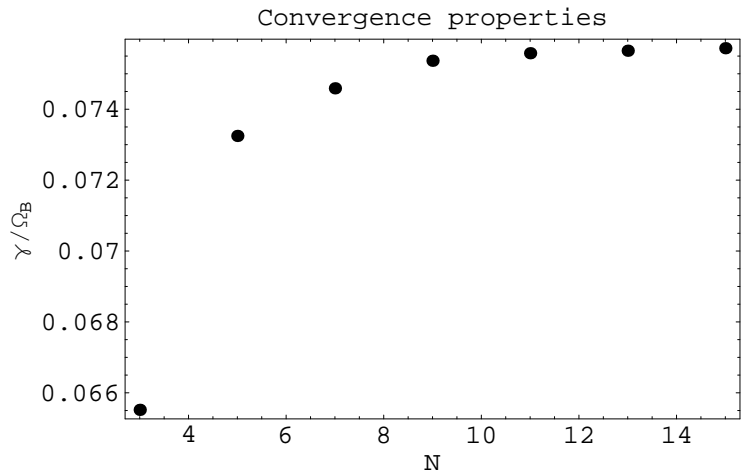}
  \caption{Example of convergence of an eigenvalue 
    when the number of terms in the expansion increases. This
    specific example corresponds to~$k_{\rm z} \, L = 0.85$ and ${\rm
      Log}_{10}\,\Theta_s= -4$ of fig~\ref{fig:Tearing}.}
  \label{fig:Convergence}
\end{figure}

\begin{figure}[htbp]
  \centering
  \includegraphics[scale=1]{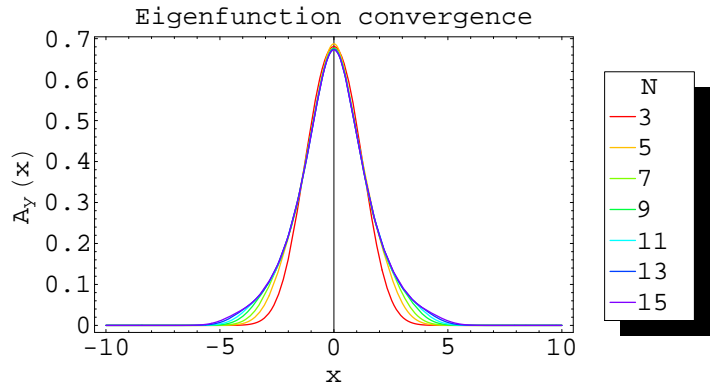}
  \caption{Example of convergence of the eigenfunction 
    associated to the eigenvalue in fig~\ref{fig:Convergence}. Only
    the coefficients associated with the even Hermite functions are
    different from zero as expected from symmetry considerations.  The
    narrowest function corresponds to $N=3$ whereas the widest to
    $N=15$.}
    \label{fig:ConvergenceFnProp}
  \end{figure}

Next, we check our algorithm against the tearing mode instabilities in
the relativistic Harris current sheet. The eigenvalue problem for the
non-relativistic Harris current sheet is given by Galeev, chapter~6.2,
page 305, in~\cite{1984bpp..book.....G}. The dispersion relation for
the non-relativistic tearing modes ($k_\mathrm{y}=0$) is given in
\cite{1991JGR....9611523P, 1995JGR...100.3551B}.  The relativistic
generalisation is given by Eq.~(\ref{eq:Harris2}).  In
figure~\ref{fig:Tearing}, we compare our numerical results with the
analytic expressions found using a simple root finding algorithm from
the dispersion relation for tearing modes in order to check the
correctness of our algorithm implementation. The growth rates are
normalised to the non-relativistic cyclotron frequency
$\hat{\Omega}_{\rm B} = q_s\,B_0/m_s$.  The numerical results in
fig~\ref{fig:Tearing} are in good agreement with the approximated
analytical expression, Eq.~(\ref{eq:Harris2}).
In the spectral decomposition method, it is necessary to choose an
appropriate number of terms~$N$ in the expansion~(\ref{eq:Her}) in
order to achieve a given accuracy. To do this, we check the
convergence properties of our method. In Fig.~\ref{fig:Convergence} we
show the eigenvalue as a function of $N$, starting with $N=3$~terms
and increasing $N$ until the eigenvalue reaches a satisfactory
precision (in this case $1\,\%$ accuracy).  The corresponding
convergence of the eigenfunction is shown in
fig~\ref{fig:ConvergenceFnProp}. Note that the results of
fig~\ref{fig:Tearing} are shown for~$N=11$.

\section{Conclusion}
\label{sec:conclusion}

We present a generalisation and extension of our previous
algorithm~\cite{2007PPCF...49..297P} to solve the linear dispersion
relation for relativistic multi-component inhomogeneous and magnetised
plasmas. The code is validated by comparing the results with two
standard configurations: the relativistic Weibel instability in a
homogeneous plasma, and the tearing mode instability in a relativistic
neutral Harris sheet.  To effect the comparison, we derived useful
analytical expressions, Eq.~(\ref{eq:RelDispWeibel}) and
Eq.~(\ref{eq:Harris2}), for the dispersion relations in these
configurations and solved them numerically.  We conclude that this
code is a suitable tool for the study of stability properties of more
general configurations of interest in gamma-ray burst and pulsar wind
theories.

\ack{This research was supported by a grant from the G.I.F., the
  German-Israeli Foundation for Scientific Research and Development.}


\end{document}